\documentclass[12pt]{article}

\usepackage[english]{babel}

\makeatletter
\usepackage{latexsym}
\usepackage[T1]{fontenc}
\usepackage{amsmath}
\usepackage{amssymb}

\def \AO {{\cal A}({\cal O})}
\def \AO' {{\cal A}({\cal O}')}

\def \] {\supseteq}

\def \limR {\lim_{R \ra \infty}}

\def \be {\begin{equation}}
\def \ee {\end{equation}}

\def \ra {\rightarrow}

\def \eqq {\equiv}

\def \b {{\beta}}

\def \eps {{\varepsilon}}

\def \om {{\omega}}

\def \A {{\cal A}}
\def \B {{\cal B}}
\def \C {{\cal C}}
\def \D {{\cal D}}
\def \F {{\cal F}}
\def \G {{\cal G}}
\def \H {\mbox{${\cal H}$}}

\def \L {{\cal L}}

\def \O {{\cal O}}

\def \T {{\cal T}}

\def \Z {{\cal Z}}

\def \id {{\bf 1 }}

\def \Psio {{\Psi_0}}

\def \x {{\bf x}}

\def \z {{\bf z}}

\def \Rbf {{\bf R}}

\def \AO {{\cal A}({\cal O})}
\def \AO' {{\cal A}({\cal O}')}

\def \] {\supseteq}

\def \limR {\lim_{R \ra \infty}}

\newfam\Ssfam
\font\eleSs=cmss10 at12pt \font\sevenSs= cmss10 at 8pt \font\sixSs= cmss10 at 6pt

\textfont\Ssfam=\eleSs\scriptfont\Ssfam=\sevenSs%
\scriptscriptfont\Ssfam=\sixSs \def\Ss{\fam\Ssfam\eleSs}

\def\doppio#1{{\rm I}\kern-.1667em{\rm #1}}

\def\Q{\text{Q}\kern-.52em
    \text{\vrule height1.5ex width.5pt depth0pt}\kern.45em}

\def\Z{{\mathchoice {\hbox{$\Ss\textstyle Z\kern-0.4em Z$}}
{\hbox{$\Ss\textstyle Z\kern-0.4em Z$}} {\hbox{$\Ss\scriptstyle Z\kern-0.25em
Z$}} {\hbox{$\Ss\scriptscriptstyle Z\kern-0.2em Z$}}}}

\def\C{{\mathchoice{\hbox{$\rm\textstyle\text{\kern.35em\vrule
   height1.5ex width.5pt depth0pt\kern-.35em C}$}}
{\hbox{$\rm\textstyle\text{\kern.35em\vrule
   height1.5ex width.5pt depth0pt\kern-.35em C}$}}
{\hbox{$\rm\scriptstyle\text{\kern.35em\vrule
   height1.5ex width.3pt depth0pt\kern-.35em C}$}}
{\hbox{$\rm\scriptscriptstyle\text{\kern.35em\vrule
   height1.5ex width.2pt depth0pt\kern-.35em C}$}}}}

\def \be{\begin{equation} \displaystyle}
\def \ee{\end{equation}}

\def \A*{\mbox{$A^{*} $}}
\def \B*{\mbox{$B^{*} $}}
\def \C*{\mbox{$C^{*} $}}

\def \id{\mbox{${\bf 1}\,$}}

\def \bea{\begin{eqnarray}}
\def \eea{\end{eqnarray}}

\def \b{\beta}

\@addtoreset{equation}{section}

\def \be {\begin{equation} \displaystyle}

\def \ee {\end{equation}}

\def \ra {\rightarrow}
\def\AO {\mbox{${\cal A}({\cal O})$}}
\def\AO'{\mbox{${\cal A}({\cal O}')$}}
\def\O {\mbox{${\cal O}$}}

\def\A{\mbox{${\cal A}$}}

\def \ra{\rightarrow}

\def \eps {{\varepsilon}}

\def \om {{\omega}}
\def \O {{\cal O}}

\def \A {{\cal A}}
\def \AO {\A(\O)}
\def \AOl'{\A(\O_{loc}')}

\def \B {{\cal B}}
\def \F {{\cal F}}
\def \D {{\cal D}}

\def \H {{\cal H}}

\def \x {{\bf x}}

\def \z {{\bf z}}

\begin{document}
\begin{titlepage}
  \title{ Local Gauss law  and local gauge symmetries \\ in QFT}

\sloppy

\author{F. Strocchi  \\  Dipartimento di Fisica, Università di Pisa, Pisa, Italy}


\fussy

\date{}

\maketitle

\begin{abstract}

Local gauge symmetries reduce to the identity on the observables, as well as on the physical states (apart from reflexes of the local gauge group topology) and therefore their use in Quantum Field Theory (QFT) asks for a  justification of their strategic role. They play an intermediate  role in deriving  the validity of  Local Gauss  Laws  {\textit {on the physical states}} (for the currents which generate the related global gauge group); conversely, we show that local gauge symmetries  arise whenever  a vacuum representation of a \textit {local field algebra} $\F$ is used for the description/construction of physical states satisfying   Local Gauss Laws, just as  global compact gauge  groups arise for the description of localizable states labeled by superselected quantum numbers. The above  relation suggests that the Gauss  operator ${\bf{G}}$, which by locality cannot vanish in $\F$, provides an intrinsic  characterization of the realizations of a gauge QFT in terms of a local field algebra $\F$ and of the related local gauge symmetries generated by ${\bf{G}}$. 

\end{abstract}



\end{titlepage} 


\section{Introduction}
 Since by definition a gauge symmetry reduces to the identity on the observables, its physical meaning has been questioned and debated from a foundational/philosophical point of view. (For  the conceptual and philosophical discussions on the empirical meaning of gauge symmetries, which appear to be relevant for our  understanding of the  physical world,  see \cite{BC}).  Clearly, the final description of a physical system should be written in a gauge invariant language and therefore gauge symmetry are bounded to play only an intermediate role.

The following physical principles help to properly set the problem of understanding the strategic role of gauge symmetries. The {\em operational/experimental description of a physical system} (not necessarily quantum!) makes reference to the following elements:

\vspace{1mm}
\noindent {\bf A}) the set  $\A$ of its measurable quantities, briefly 
called {\bf observables};

\vspace{1mm}
\noindent {\bf B}) the measurement of the {\bf time evolution} of the observables;

\vspace{1mm}
\noindent {\bf C}) the set $\Sigma$ of configurations or {\bf states}  in which the system may be prepared, according to well defined protocols of experimental preparations; the measurement of an observable $A$, when the system is in the state $\om$, is  operationally defined by the {\bf experimental expectation} $< A  >_\omega$.

Given a state $\om$, the set of states  which can be prepared starting from $\om$, through physically realizable operations, is denoted by $\Gamma_\om$ and called the {\bf phase} of $\om$. This means that the protocols of preparations of states belonging to different phases are not related by physically realizable operations.
Clearly, by definition, different phases describe disjoint realizations of the system (or disjoint ''worlds'') which cannot communicate through realizable  observable operations.

The above \textit{operational framework} has an essentially unique \textit{mathematical  transcription}.
General physical considerations indicate that the set of the observables generate a normed algebra, actually a $C^*$-algebra with identity, called the {\bf algebra of observables}, for simplicity still denoted by $\A$. The time evolution defines a one-parameter  group of transformations (technically automorphisms) of $\A$: $t: \A \ra \alpha_t(\A)$.

 By  its experimental expectations, each state $\om$, defines a linear positive functional  $\om(A) \eqq <A>_\om, \, \forall A \in \A$, and, by the GNS theorem, a representation $\pi_\om$ of $\A$ (as operators $\pi_\om(A)$) in a Hilbert space $\H_\om$; $\om$ is represented by a vector $\Psi_\om \in \H_\om$ and the experimental expectations are represented by the matrix elements $<A>_\om = (\Psi_\om, \pi_\om(A) \, \Psi_\om)$. (For a  discussion of the general description of a physical system, see \cite{FS12}). 

The  set of states  $\pi_\omega(\A) \, \Psi_\om$, are dense in $\H_\om$, from  which one may obtain also  the mixed states associated to $\pi_\om$. The set of all such states is called the folium associated to $\om$, which  may be considered as the mathematical description of $\Gamma_\om$. The family of  physical states on $\A$ is denoted by $\Sigma(\A)$.

In the following, we shall consider infinitely extended systems, which generically display  the occurrence of different phases and the related role of gauge symmetries (see below).  For this kind of systems a very relevant  property  is the {\bf local structure} of $\A$: each  observable   is identified by the experimental apparatus used for its measurement and, since the physically realizable operations, as well as the corresponding experimental apparatuses, are inevitably localized in space, the algebra of observables is generated  by  localized observables .  This means that the algebras $\A(V)$ of observables localized in bounded regions  $V$, generate $\A$,  $\A = \cup_V \,\A(V)$, (hereafter called local algebra). 

Another relevant property is the stability  of $\A$ under the group of {\bf space translations}  $\alpha_{\bf a}, \,\,{\bf a } \in \Rbf^3$, with  $\alpha_{\bf a}(A)$ denoting the ${\bf a} $-translated of $A$. (For lattice systems the space translations are replaced by the lattice translations). 

  For a reasonable physical interpretation,
 the  measurement  of a localized observable $A$ should not  be influenced by measurements of  observables at infinite space separations, i.e. the following condition (called  {\bf asymptotic abelianess} must hold, ($B$ a localized observable):
  $\lim_{|\x| \ra \infty}\, [\,A, \,\alpha_\x(B)\,] = 0.$

A distinguished role is played by the \textit {pure homogeneous states} $\om_0$, characterized  by being \textit{invariant under a subgroup $\T$ of translations}
\be{ \om_0( \alpha_{\bf a} (A)) = \om_0(A), \,\,\,\forall A \in \A}\ee
and by satisfying the {\em cluster property}\index{cluster property}
\be{ \lim_{n \ra \infty}\,\om_0(A\,\alpha_{n \,{\bf a}}(B)) = \om_0(A)\,\om_0(B),}\ee
where $n \,{\bf a}$ denote the group parameters of $\T$. Typically, but not necessarily, $\om_0$ is a ground state. The set of such states $\om_0$ on $\A$ is denoted by $\Sigma_0(\A)$. 

A very important consequence  is that:
 
\noindent $\om_0$ {\em is the unique $\T$ invariant state in the pure homogeneous phase $\Gamma_{\om_0}$}.

The interest of considering the phases  defined by such states  is twofold. 

In relativistic quantum field theory, the vacuum state is  invariant under space translations and the validity of the cluster property is a necessary condition for the possibility of defining the scattering matrix, which requires the factorization of expectations of infinitely (space) separated clusters (describing scattering processes). Thus, the phase defined by a vacuum state in quantum field theory is a homogeneous pure phase in the above sense. 

Quite generally,  in a homogeneous pure phase  the macroscopic observables, defined by space averages of local observables, take sharp (classical) values in agreement with the characteristic property of the standard pure phases in thermodynamics. 

In conclusion, general physical principles characterize the mathematical description of a physical system, in terms of the local algebra of observables $\A$, its time evolution $\alpha_t(\A), t \in \Rbf$,  and the family of states $\Sigma(\A)$ in which the system may be prepared. The task of theoretical physics is to devise concrete effective strategies for implementing  and controlling such a general structure by  determining    the representations of the observable algebra corresponding to the various (physically realizable) states. 

\def \Fo  {\F_{obs}}

In quantum field theory, as advocated by Wightman, the standard and well established  strategy is to study the vacuum representations of the relevant  field algebra, provided by the vacuum correlation functions. However, already in the case of a system of free particles,  physically important  particle states are not present in the vacuum representation of the observable algebra, called the vacuum sector. 

A simple example  is provided by the system of free massive Dirac fermions, since the observable algebra and therefore its vacuum representation have zero fermionic charge ;  a complete  QFT description is obtained by introducing the  local field algebra $\F$ generated by the fermionic fields, whose (unique) vacuum representation contains  the representations of the  observable sub-algebra  $\Fo$, labeled by the quantum number of the fermionic charge.   

This is  the standard strategy  adopted in the  formulation  and control of QFT models. One studies the vacuum representation of a local field algebra $\F$ generated by local fields, which describe the degrees of freedom inferred and borrowed  from the limit of vanishing interaction. Such a local field algebra contains a (suitably characterized) observable field sub-algebra $\Fo$ and  one looks  for the representations of $\Fo$ contained in a vacuum representation of   $\F$. 

Therefore, the assumed role of the field algebra $\F$ is to provide the building stones of the theory: the fields of $\F$ describe the relevant states,  generate the observable algebra and allow for a direct and simple definition of the dynamics (through a Hamiltonian or Lagrangian which is a polynomial function of the fields of $\F$).

 In conclusion,  the difficult problem of determining the representations of the local observable algebra and its time evolution,  beyond the vacuum sectors, i.e. the solution of the problem $(\A, \alpha_t(\A), \Sigma(\A))$, is attacked  by finding the vacuum representations of a local field algebra $\F \supset \A$, i.e. by solving the relatively easier problem $(\F, \alpha_t(\F), \Sigma_0(\F))$.

Behind  such a strategic  choice there is  the implicit assumption (extrapolated from the non-interacting case) of considering  the states outside the vacuum sector which  are local states, i.e. may be obtained by applying the local fields of $\F$ to the vacuum; this property allows for their localizability in the DHR sense (for general discussion of such an important property, see \cite{Ha}):  the state $\om$ is localized in the double cone  $\O$ if 
\be{ \om(A) = \om_0(A), \,\,\,\,\, \forall A \in \A(\O'), }\ee
where $\O'$ denote the set of points which are spacelike  w.r.t.   $\,O$ and $\A(\O')$ the algebra of observables localized in $\O'$.

The quantum numbers which characterize the  inequivalent representations of $\A$ contained  in the Hilbert space $\H_{\om_0(\F)}$  of a vacuum representation of $\F$, define superselection rules; they   identify  a global gauge group $G$, which  defines non-trivial symmetries  of  $\F$ and   it is represented by the states of $\H_{\om_0(\F)}$ (if $\om_0(\F)$ is invariant under $G$). 

Under general assumptions (namely the absence of infinite degeneracy  of particle types with equal mass and the completeness of the asymptotic states) the so derived global gauge group is compact. These deep results, which explain the {\textit {origin of compact gauge groups  from general physical principles}} were obtained in a series of papers by Doplicher, Haag and Roberts. (For a comprehensive  account see \cite{Ha}, esp. Section IV.4, and references therein.) 

	In conclusion, the experimentally detectable existence of \textit{localizable states labeled by superselected quantum numbers}  imply that (under general assumptions) a local field algebra which generates them from the vacuum  has the symmetry of a global compact gauge group; this explains why such gauge groups  arise in the theory of infinitely extended systems, typically in quantum field theory, where locality plays a crucial role and vacuum representations of  local field algebras are the object of the standard approach.  

More subtle is the problem of justifying  the role of local gauge symmetries on the basis of general physical principles, since they reduce to the identity both on the observables and on the (physical) states. There is no doubt that local gauge symmetries proved to be  useful  for the formulation of the quantum field theory of elementary particles, but an interesting question is whether their \textit{raison d'ètre}  may be traced back to general a priori arguments beyond their a posteriori successful use.

The standard justification of local gauge symmetries, in particular   in the prototypical case of quantum electrodynamics,  is tight to the introduction of redundant degrees of freedom through the vector potential, used either  for describing a massless spin one particle (the photon) or for  defining  the electron-photon interaction.\cite{We} \goodbreak

These reasons do not appear to be  rooted in general principles better than the origin of local gauge symmetries they should explain.
Actually, the photon is equally well described by the electromagnetic (quantum) field $F_{\mu\, \nu}$, whose Lorentz transformation does not require a gauge transformation. 
As a matter of fact,  already for the solution of the  free field equations $\partial^\mu F_{\mu\, \nu} = 0$, the introduction of the quantum field $A_\mu$ as a Hilbert space operator is not free of problematic choices, since it cannot be done    without giving up covariance and/or locality (see \cite{FS16}, Chapter 7, Appendix 8).

   Moreover, a Lagrangian invariant under the group $\G$ of  local gauge transformations, labeled by the set of infinitely differentiable  gauge functions \textit{localized in space}, does not  define a deterministic time evolution of the field algebra and therefore $\G$ must be broken by a gauge fixing; such a breaking need not to be down to the identity, as might be inferred from the functional integral argument,  but only up to the extent of restoring  deterministic evolution  (see  \cite{FS19}). 

In the standard  approach to gauge theories (particularly in the perturbative treatment) the identification and construction of the physical states is usually obtained by starting with a Lagrangian invariant under a group  $\G$ of local gauge transformations and  by adding a gauge fixing, which preserves the invariance  under the related  global gauge group $G$ and allows to use a local field algebra $\F$. Then, the (infinitesimal) transformations of $G$ on $\F$ may be  generated by local currents  $J^a_\mu$, ($a = 1, ...n,  \,n  =$ the dimension of $G$). 

As a consequence of the  gauge fixing, the second Noether theorem does not apply  and such currents do not satisfy a Local Gauss Law, i.e. they are not  the divergence of  antisymmetric tensors $ F^a_{\nu\,\mu}$. Nevertheless,  one may show that a characteristic property of the physical states constructed in a vacuum representation of $\F$ is that the currents $J^a_\mu$ satisfy Local Gauss laws  on them \cite{FS16, FS19}; i.e.  
for any physical state $\Psi \in \H_{\om_0(\F)}$
\be{ (\Psi, (J_\mu^a - \partial^\nu F^a_{\nu\,\mu})\, \Psi) = 0,}\ee
(with $F^a_{\mu\,\nu}$ actually the field  strength) or, in a manifestly gauge invariant way, 
\be{(\Psi, \sum_a (G^a_\mu)^*\, G^a_\nu \,\Psi) = 0, \,\,\,\,\,\,\,\,G^a_\mu \eqq J_\mu^a - \partial^\nu F^a_{\nu\,\mu}.}\ee
Such a form of the Local Gauss Law  is obviously satisfied in QED and may be easily checked to hold in the (local) BRST quantization of Yang-Mills theories, thanks to the nillpotency of the BRST charge. Actually, for the same reasons, any monomial  of  $G^a_\mu$ has vanishing expectation on the physical states. For brevity, such physical states  shall be called \textit{LGL states}. Clearly a prototypical  example is provided by the charged states in QED.

Thus, one may argue that even if local gauge symmetries reduce to the identity on the physical states, they play an intermediate role in the construction of the representations of the observable algebra by guaranteeing that they are defined by  physical states obeying LGL.

Conversely, given the existence of LGL states,  one may investigate which gauge symmetries emerge for a local field algebra $\F$ which allows for their construction through its vacuum representation; we shall argue that in this way one obtains local gauge symmetries.  This means that, in order to allow for the construction of LGL states, the local algebra  $\F$ must contains fields with non-trivial transformation under  a local gauge symmetry.

This offers a possible explanation of the emerge and strategic role of local gauge symmetries,  \textit{without  ever mentioning the vector potential}, which, being generically determined up to a (local) gauge transformation, automatically and obviously brings with it the freedom of  local gauge symmetries. 

In this perspective local gauge symmetries are not introduced by an \textit{a priori} ansatz or Local Gauge Principle, nor as a consequence of the introduction of  redundant degrees of freedom through the vector potential, but automatically arise as symmetries  of a local field algebra which  allows for  the realization of states satisfying Local Gauss Laws.

\section{Local Gauss Laws and local gauge symmetries}

The aim of this Section is to argue that  physical LGL states (e.g. the charged states in QED or  unconfined quark states in QCD) lead to the emergence of local gauge symmetries if a   local field algebra is  used for their construction.

In the abelian case (QED)  the characterization of the  physical LGL states is simply provided by the electrically charged  states, labeled by the superselected electric charge (giving rise to a $U(1)$ global gauge group) and satisfying the  LGL  given by the Maxwell equations.

 Less obvious  is the non-abelian case and we adopt the following characteristic properties of LGL states, extracted from their actual construction:

\noindent i) they carry superselected quantum numbers corresponding to a compact global gauge group $G$, as in the DHR case,

\noindent ii) in contrast with the DHR states, LGL states cannot  be described by local states in a vacuum representation of an auxiliary  local field algebra $\F$ which contains the field strengths $F_{\mu\, \nu}^a$, as well as the  currents $J^a_\mu$, which generate the infinitesimal transformations of $\F$ under $G$; for brevity,  the corresponding charges shall  be called \textit{Gauss charges}, to emphasize  that as a consequence of the LGL they are not localizable charges,

\noindent iii) in the abelian case of Quantum Electrodynamics (QED) they do not define local states on the algebra of the observables and therefore they  are not localizable in the DHR sense.

\vspace{1mm}
\noindent {\bf Remark 1}.\,\,
Property ii) follows from the fulfillment of LGL, if the vacuum representation of $\F$ satisfies semipositivity or the relativistic spectral condition.
In fact, eq. (1.5) with $\Psi = \Psio$, implies that $ G^a_\nu \,\Psio$ is a null vector and therefore, if  semipositivity holds,  $(\Psi, G^a_\nu\,\Psio) = 0$, $\forall \Psi \in  \H_{\om_0(\F)}$. As a consequence of this last equation, given  a local state $\Psi = F\,\Psio$, where $F \in \F$ transforms non-trivially under $G$, 
\be{ \delta^a F = i  \limR [ \,Q^a_R, \, F\,],   \,\,\,\,\,Q^a_R = J^a_0(f_R\, \alpha),}\ee
(with the standard notation for the smearing of $J^a_0$,  see e.g. \cite{FS16}, Chapter 7, Section 2), by the locality of $F$ one has 
$$ 0 \neq (\delta^a F\,\Psio, \delta^a F \Psio) = i \limR (\delta^a F \Psio,  [\,J^a_0(f_R \alpha) - \partial^i F_{i\, 0}(f_R \alpha), \,F\,] \Psio).$$ Then, if $F \Psio$, and therefore $\delta^a F \Psio$, satisfies LGL, the r.h.s. reduces to    $$ i \limR ( F^*\,\delta^a F \Psio,  G^a_0(f_R \alpha)\, \Psio) = 0, $$
leading to  a contradiction. 

The same conclusion is reached even if the vacuum correlation functions of the auxiliary local field algebra $\F$ do not satisfy semipositivity, but the relativistic spectral condition   holds. In fact, as above, by locality for a local state $\Psi = F\,\Psio$ one has 
$$ \limR < \Psi, [\, Q^a_R, F\,] \Psio> =  \limR < \Psi, [\,G^a_R, F\,] \Psio>,  $$
the limit is reached for finite $R$ and, if $\Psi $ satisfies   the LGL, the r.h.s reduces to $< \Psio, F^* \,F \,G^a_R \,\Psio >$, $R$ large enough. 

Now, by the relativistic spectral condition  $$W(y-x, z-x) \eqq < \Psio, F^*(x) \,F(y) \,G^a_R(z) \,\Psio >,$$ (where $F(y) \eqq U(y) F  U(y)^{-1} $ denotes the $y$-translated of $F$, and similarly for the other operators), is the boundary value of a function $W(\zeta_1, \zeta_2)$ analytic in the tube $\T_2$. Such  a function vanishes for  $\zeta_1, \,\zeta_2$ real, $\zeta_2 = (\z_2, z_{0, 2})$, $|\z_2|$ sufficiently large, since then   $[ \,F(y), \,G^a_R(z)\,] = 0$ and LGL applies. Then, by the edge of the wedge theorem $W(\zeta_1, \zeta_2) = 0$ and $F \,\Psio$ is chargeless, contradicting i).

\vspace{1mm}
\noindent {\bf Remark 2}.\,\,In the QED case, the current, which generates the global  gauge group $U(1)$, and the field strength $F_{\mu\, \nu}$ are observable fields; then, if  a state  $\om$ is localized in the DHR sense, say in a double cone $\O$, for sufficiently large $R$, $\partial^i F_{0 \, i}(f_R \alpha) \in \F_{obs}(\O')$, so that, according to the DHR criterion,    $\om(\partial^i F_{0 \, i}(f_R \alpha) ) = \om_0(\partial^i F_{0 \, i}(f_R \alpha)) =0$ and, by the LGL, $\om$ is chargeless. 

\vspace{2mm}
The embedding of the observable algebra into a larger algebra $\F$ for the description of LGL states, through  a vacuum representation of $\F$, is not unique and the local gauge symmetries of $\F$ depend on $\F$. As we shall see, the generators of the infinitesimal  local gauge transformations on $\F$ are  provided by the Gauss operators $G^a_0$.    A limiting case in QED is the choice of the Coulomb gauge field algebra $\F_C$, since in $\F_C$ the Gauss operator $G_0$ vanishes, $\F_C$ is non-local  and  the Coulomb  gauge fixing excludes any local gauge symmetry of $\F_C$, (for the general structure of the Coulomb gauge see \cite{Di, Sy}. For the necessary ultraviolet regularization, see, for a perturbative control, \cite{Ste} and, for a general control which exploits the properties of the Feynman-Gupta-Bleuler (FGB) gauge, \cite{BDMRS}; see also the discussion in \cite{FS19} Section 2.3).    

Actually,  a crucial property for the strategy outlined above, leading to local gauge symmetries,  is the locality of the field algebra; this choice, motivated by the no-interaction limit,  is also suggested by technical reasons, since locality helps for the control of the dynamics (it is well known that the renormalizable gauges are local gauges). 
Also from a constructive point of view, the infinite volume limit is better handled  for a local field algebra with a local dynamics.\footnote{For the problems arising for non-local dynamics, see e.g. \cite{FS21}, Appendix A. 

\noindent  In the case of DHR states the existence of a local field algebra for their description is guaranteed by general principles; in the case of LGL states it may be motivated by the standard way of treating gauge field theory models, e.g. in the perturbative treatment of QED or more generally of the standard model. Some hint is provided, e.g. in 	QED, by giving a small mass $\mu$ to the photon, so that the charged states become  DHR states and DRH analysis applies with the existence of  a local field algebra $\F$ for their description. As shown by Blanchard and Seneor the vacuum correlations of $\F$ have a  limit for $\mu \ra 0$ (preserving locality) and the problem is reduced to the construction of the physical LGL states in terms of the vacuum representation of the local algebra $\F$. 
Actually, in Symanzik's  treatment of the Proca theory, with the use of the Stuckelberg field $B$, in analogy with 	QED, (\textit{Lectures on Lagrangian Field Theory}, DESY report  T-71/1), the fields $\psi_g = \exp^{-ie [ (- \Delta)^{-1 } \partial_i A^i]} \psi$, \,$A_g^\mu = A^\mu - \partial^\mu [(-\Delta)^{-1} \partial_i A^i]$ , with $\psi, \, A_\mu$ the Proca fields, commute with $B$, which in the limit $\mu \ra 0$ generates the local gauge transformations, (with gauge  parameters $\eps(x)$ satisfying $\square \eps = 0$), and should yield  the Coulomb charged states.}  

For these reasons,  the auxiliary local field algebra $\F$  should not satisfy the LGL; otherwise, by locality, $\F$ would be pointwise invariant under the global gauge group $G$. \goodbreak

In fact, if the LGL hold in the local algebra $\F$ one would have
$$ \delta^a F = \limR [\, J^a_0(f_R \alpha), \, F\,] = \limR [\, \partial^i F^a_{i \,0}(f_R \alpha), \, F\,] = 0,\,\,\,\,\,\forall F \in \F, $$
 and, consequently,  $\H_{\om_0(\F)} = \overline{\F \, \Psio}$ would not contain states carrying  non-trivial charges of $G$.

   The different choices of the local algebra $\F$ are characterized by the way LGL fail, i.e. by the non-vanishing Gauss operators $G^a_0$. To this purpose, it is useful to remark that LGL correspond to a combination of hyperbolic evolution equations and constraint/elliptic  equations for the field strengths $ F^a_{\mu \,\nu}$:
\be{ \square F^a_{\mu \,\nu} = \partial_\mu J^a_\nu - \partial_\nu J^a_\mu + C^a_{\mu\,\nu}, \,\,\,\,\,\,\,\,\,\,\,\,\,\,\partial^i F^a_{i \,0} = J^a_0,}\ee
with $C^a_{\mu\,\nu}$ a bilinear function $A^c_{\lambda}, F^b_{\rho\,\sigma}$ and their first derivates.    If, as required, LGL do not hold in $\F$, the above equations get modified by the non-vanishing Gauss operators   $G^a_\mu$,\,\, $ \partial^\mu G^a_\mu =0$.

\vspace{1mm}
\noindent a) {\bf{ \textit{Time independent local gauge symmetries}}}

As remarked above, a non-trivial representation of the global gauge group $G$ by a local field algebra $\F$  requires that $G^a_0 \neq 0$, since, by locality  the infinitesimal transformations of $\F$  by $G$ are given by  
$$\delta^a F = \limR [\, J^a_0(f_R \alpha), \, F\,] = \limR [\,J^a_0(f_R \alpha) -  \partial^i F^a_{i \,0}(f_R \alpha), \, F\,] =$$
\be{ =  \limR [\, G^a_0(f_R \alpha), \, F\,].}\ee
 A  possible realization of such a framework,  which, in a certain sense, minimizes the needed violation of the LGL in $\F$, even at the expense of loosing manifest covariance, is given by  $G^a_0 \neq 0$, 
$ G^a_i  = 0$, which plays the role of a gauge fixing; then, the continuity equation for $G^a_\mu$ requires  that $G^a_0(\x, t)$  is time independent. 
As a consequence, eqs.\,(2.2) are replaced by 
\be{ \square F^a_{i \,j} = \partial_i J^a_j - \partial_j J^a_i + C^a_{i \,j},\,\,\,\,\,\,\,\,\,\,\partial^0 F^a_{0\, i} =  - \partial^j F^a_{j \, i} + J^a_i,}\ee
with the equal time constraint 
\be{\partial^i F^a_{i\, 0 } =   J^a_0 - G^a_0.}\ee
This is the choice adopted by the temporal gauge; in fact, eqs.\,(2.4, 2.5) are the evolution equations of the fields $F^a_{\mu \,\nu}$ in the temporal gauge. 

Equations (1.5) imply that  for any physical state $\Psi$, $(G^a_0 \,\Psi, \,G^a_0\, \Psi) = 0$, i.e., if positivity holds,
$G^a_0\, \Psi = 0$. 

One should remark that a mathematical subtlety occurs in the standard local and positive   temporal gauge, namely    the local field algebra $\F$ represented by a vacuum state  is generated by the exponentials  of the standard (non-observable) fields, with algebraic relations corresponding to those of their formal generators. Then, the condition which selects  the physical states   should rather read $$V^a(\Lambda) \Psi = \Psi,\,\,\,\,\,	\Lambda  \in \D(\Rbf^3),$$
where  the unitary operator  $ V^a(\Lambda) $ is formally the unitary  exponential of $G^a_0(\Lambda)$.\goodbreak

Furthermore, since $\F$  is not covariant under  relativistic transformations the relativistic spectral condition is not satisfied by the vacuum correlation function of $\F$, the Reeh-Schlieder theorem does not apply and $G^a_0 \Psio = 0$, or better $(V^a(\Lambda) - \id) \, \Psio = 0$ does not imply that the local operator $G^a_0$ or better $(V^a(\Lambda) - \id)$ vanishes.\footnote{ For a more detailed discussion see  \cite{FS16} Chapter 8, Section 2.1.}

 For simplicity, in the following discussion, we shall sometimes use  the formal generators of the exponential fields, the more accurate mathematical discussion being easy to obtain.

The operator $G^a_0$ plays the role of a static charge density which is not seen by the physical states and compensates the vanishing flux of $ F^a_{i \,0}$  at  (space) infinity on the local states of $\H_{\om_0(\F)}$.

The non-vanishing  $G^a_0$ modifies the equal time constraint implied by the LGL on the local states; the  vanishing of $G^a_0$ on the physical states requires that the physical states of $\H_{\om_0(\F)}$, carrying a non-trivial charge of $G$, are non-local limits of local states. 

Now, for any test function $\Lambda(\x) \in \D(\Rbf^3)$, the operator $G^a_0(\Lambda, t) $ generates a \textit{time independent derivation} on $\F$ 
\be{ \delta^{a, \Lambda} F \eqq i [\, G^a_0(\Lambda, t), \,F\,] = i [\, G^a_0(\Lambda, h), \,F\,], \,\,\,\,\,\,h \in \D(\Rbf), \,\,\,\,\,\int d t\,h(t) = 1. }\ee
This is the basic property of the derivations generated by local (covariant) currents, corresponding to infinitesimal symmetry transformations, where the time independence is a consequence of current conservation and locality. 

If the auxiliary local field algebra $\F$ contains the formal exponentials of $G^a_0$, (as in the standard temporal gauge), the following local transformations are defined on $\F$:
 \be{ \b^\Lambda(F) = V^a(\Lambda)\, F \,V^a(\Lambda)^*, \,\,\,\,\,\, \forall F \in \F.}\ee    

Moreover, since  the subspace of physical state vectors  must be pointwise invariant under the application of the observable operators, these operators  should commute  with $V^a(\Lambda)$, so that the above  transformation defines a \textit{local time independent gauge transformation}.

As a matter of fact, in the temporal gauge, where  the local algebra $\F$ is generated by canonical fields, the derivation (2.6) corresponds to  the standard time  independent gauge transformations, with local gauge parameter $\Lambda(\x)$. 

The time independence of the generators $G^a_0$ implies that the time evolution of $\F$ commutes with such local gauge transformations; thus, the Hamiltonian, as a function of the fields of $\F$, must  be gauge invariant, a property which in the standard approach corresponds to the requirement of minimal coupling.

Thus, by the above arguments,  the strategic role  of local gauge symmetries may be traced back  to the realization of LGL states through a vacuum representation of a local field algebra $\F$, where the necessarily \textit{non-vanishing  Gauss operators  generate  local gauge symmetries}.\goodbreak

\vspace{3mm}
\noindent b) {\bf \textit{Local gauge symmetries in QED}}

Another distinguished example of local gauge symmetries, arising according to the pattern discussed above, is provided by Quantum Electrodynamics.\footnote{For a general discussion of the occurrence  of local gauge symmetries in QED, in a  C*-algebra setting, see F. Ciolli, G. Rizzi, E. Vasselli, QED Representation for the Net of Causal Loops, Rev. Math. Phys., {\bf{27}}, 1550012 (2013);   D. Buchholz, F. Ciolli, G. Rizzi, E. Vasselli, The universal algebra of the electromagnetic field. III. Static charges and emergence of gauge fields, arXiv: 2111.01538 [math-ph]. When the present note was in reparation a very important result was obtained by the same authors, for QED in the presence of external charges \cite{BCRV}. } 

This example is obtained by requiring that the time evolution of the observable  electromagnetic field as operator in $\F$ is not modified by the non-vanishing Gauss operator $G_\mu$, i.e. that 
\be{  \square F_{\mu \,\nu} = \partial_\mu J_\nu - \partial_\nu J_\mu,}\ee
the only effect of $G_\mu \neq 0$ being a modification of the equal time constraint
\be{\partial^i F_{i \,0} = J_0 - G_0.}\ee
This implies that $\partial_\mu G_\nu - \partial_\nu G_\mu = 0$ and in a \textit{Lorentz-covariant local field algebra} this is obtained by  $G_\mu  = \partial_\mu \L$, with $\L(x)$ a scalar field of $\F$; then, the continuity equation obeyed by $G_\mu$ implies  $\square \L(x) = 0$. 

Such a choice of the local field algebra corresponds to the  Feynman-Gupta-Bleuler (FGB) gauge, albeit in a more general context, since no reference is made to the vector potential; in fact,   eqs.(2.8), (2.9) coincides with the equations for $F_{\mu\, \nu}$ in  that gauge. 

The relativistic covariance of $\F$ lead to  the validity of the Reeh-Schlieder theorem and therefore positivity cannot hold, since otherwise eq.\,(1.5) implies $G_\mu \,\Psio = 0$; hence, by the Reeh-Schlieder theorem the local operator  $G_\mu$ vanishes, $\F$ commutes with the charge and $\H_{\omega_0(F)}$ does not contain charged states. 

 A way out of this difficulty is to guarantee the validity of eq.\,(1.5) on the vacuum state by a non-local condition; since $\L(x)$ is a free field, its negative energy part $\L(x)^-$ is well defined, it is a non-local operator and  the equation
\be{ \partial_\mu \L(x)^- \, \Psi   = 0}\ee
implies that $\Psi$ satisfies eq.\,(1.5); then, such a condition may  be chosen for  selecting the physical states.

As in the temporal gauge discussed above for the general case of a compact gauge group $G$, one may show that in QED the Gauss operator generates a time independent derivation on the local field algebra $\F$.

To this purpose, one considers  infinitely differentiable functions $\Lambda(x)$ satisfying
\be{ \square \Lambda(x) = 0,  \,\,\,\,\,\, \,\,\,\,\,\,\,\Lambda(\x, 0),\,\,\partial_0 \Lambda (\x, 0) \in \D(\Rbf^3), }\ee
and  the operator 
\be{ G(\Lambda)_{R}(x_0) \eqq   \int d^3 x \,f_R(\x)\,( \Lambda(x) \stackrel{\leftrightarrow}{ \partial _0} \L(x)),}\ee
 with $(A \stackrel{\leftrightarrow}{ \partial _0} B) \eqq (A \partial_0 B - \partial_0 A \,B)$.

Since  $\Lambda(x) $ and  $\L(x)$ satisfy the free wave equation, by  locality the following commutator   is independent of time, i.e.
\be{ i \partial_0 \limR [ G(\Lambda)_{R}(x_0), \, F] = 0,\,\,\,\,F \in \F.}\ee
In fact,
$$\partial_0 \limR [ G(\Lambda)_{R}(x_0) =   \int d^3 x\, f_R(\x) (\Lambda(x) \Delta \L(x) - \Delta \Lambda(x)  \L(x)) = $$
$$ =  \int d^3 x\, \partial_i f_R(\x) (\Lambda(x) \partial_i \L(x) - \partial_i \Lambda(x)  \L(x)).$$
Then, since supp $\partial_i f_R(\x) \subset (R \leq |\x| \leq R(1 + \eps))$, by locality the commutator (2.11) vanishes.
In conclusion, one has  a time independent derivation on $\F$, labeled by the  infinitely differentiable functions $\Lambda(x)$ of compact support in space
\be{ \delta^\Lambda \,F \eqq \int d x_0\, \alpha(x_0) i \limR \,[\, G(\Lambda)_R(x_0), \, F\,].}\ee

The stability of the subspace of physical states under application of observable operators is guaranteed if the observables commute with $\L(x)^-$, (and therefore  with $\L(x)$). Then the above derivation has the meaning of  an infinitesimal local gauge transformation with gauge function $\Lambda(x)$. 
Indeed, if the algebra $\F$ is generated by local canonical fields, as it is the case of the FGB realization, one gets the standard local gauge symmetries   of the FGB gauge. 

As discussed above in point a), the time independence  of the derivation (2.14) implies that, as a   function of the local fields of $\F$, the Hamiltonian should be invariant under local gauge transformations with gauge parameter $\Lambda(x)$, (satisfying $\square \Lambda = 0$).

\section{Conclusion}

In conclusion, a possible physical/empirical explanation of gauge quantum field theories is the existence of (particle) \textit{states which carry the quantum numbers of a  (global) compact  gauge group $G$  and satisfy Local Gauss Laws}, for the conserved currents associated to $G$. The prototype is clearly  Quantum Electrodynamics where the electric charge (the generator of the global compact $U(1)$ gauge group)  labels  the (physical) charged states, which  satisfy the  Local Gauss Law corresponding to the Maxwell equations.  Quite generally, the  description of such  LGL states through a vacuum representation of a {\bf \textit{local}} \textit{field algebra} $\F$  leads to the emergence  of {\bf \textit{local}} \textit{ gauge symmetries} for $\F$, which commute with the time evolution of $\F$. 

 Thus, whereas   states carrying localizable superselected charges lead to global compact  gauge  groups  for  the local field algebra which obtains them from the vacuum,  the realization of  states carrying  Gauss charges through a  vacuum representation of a local field algebra $\F$ implies a  local gauge symmetry for $\F$,  i.e. such $\F$  must contains fields which transform non-trivially under local gauge transformations. Therefore, the local extension of the global gauge group  $G$ (related to the superselected quantum numbers) needs not to be \textit{ a priori} assumed with no compelling physical motivations (as in the standard   characterization/definition of  gauge theories),  but it is required by the \textit{physical} existence  of states carrying Gauss charges and their construction through the vacuum representation of a local field algebra. 

     In our opinion, the role   of the Gauss operator as the generator of the local gauge symmetries suggests  a better classification of the possible  local field algebras used for realizing the LGL states, better than the gauge fixings which  typically involve the vector potential.  

\vspace{10mm}
\noindent  
Acknowlegements.
I am indebted to Francesco Serra for a stimulating discussion.


\end{document}